\documentclass[twocolumn,superscriptaddress,prb,showpacs]{revtex4-1}

\usepackage{amsmath,amssymb}
\usepackage[dvipdfmx]{graphicx}
\usepackage{bm}
\usepackage{times,txfonts}

\allowdisplaybreaks

\begin{document}

\title{
Crossed responses of spin and orbital magnetism in topological insulators
}
\date{\today}
\author{Ryota Nakai}
\affiliation{WPI-Advanced Institute for Materials Research (WPI-AIMR), Tohoku University,
Sendai 980-8577, Japan}
\email{rnakai@wpi-aimr.tohoku.ac.jp}
\author{Kentaro Nomura}
\affiliation{Institute for Materials Research,
Tohoku University, Sendai 980-8577, Japan}

\begin{abstract}
Crossed magnetic responses between spin and orbital angular momentum are studied in time-reversal symmetric topological insulators.
Due to spin-orbit coupling in the quantum spin Hall systems and three-dimensional topological insulators, the magnetic susceptibility has crossed (intersectional) components between spin and orbital part of magnetism.
In this study, the crossed susceptibility for the orbital magnetization is studied in two- and three-dimensional topological insulator models, in which an external magnetic field interacts with the electron spin by Zeeman coupling via distinct g-factors for conduction and valence energy bands.
The crossed susceptibility in two-dimensional quantum spin Hall insulators shows a quantized signature of the $\mathbb{Z}_2$ topological phase in response to Zeeman coupling via an averaged g-factor, and the quantization persists even when $\sigma^z$-conservation of electrons is broken by a tilted magnetic field.
The bulk orbital magnetization is interpreted by the persistent edge current attributed to the chiral anomaly at the (1+1)-dimensional boundary.
In three-dimensional topological insulators, we found that the crossed susceptibility is proportional to the difference of g-factors of conduction and valence electrons, which is qualitatively different from the two-dimensional case.
Steep changes of the crossed susceptibility in three dimensions at the phase transition points are explained by the surface Dirac fermion theory.
Finally, dependence of the crossed susceptibility on g-factors in two- and three-dimensional cases is discussed from the viewpoint of time-reversal and particle-hole symmetries.
\end{abstract}


\maketitle

\section{Introduction}

Characteristic response phenomena are signature of topological phases of matter.
The quantum Hall insulators show quantization of the electronic Hall conductivity in unit of $e^2/h$,
and its quantization number counts the Chern number of occupied electronic energy bands\cite{prange87,thouless82}.
As for thermal responses, quantum Hall insulators and two-dimensional topological superconductors show the quantized thermal Hall conductivity in units of $k_{\text{B}}^2T/3h$ and $k_{\text{B}}^2T/6h$ \cite{kane97,read00}.
In both cases, Hall conductivities are nonzero when time-reversal symmetry is broken, since the time-reversal operation inverts the sign of those Hall conductivities.

The quantum Hall effect is efficiently described in terms of the thermodynamic quantities by the St\v{r}eda formula\cite{streda82}
\begin{align}
 \sigma_{H}
  =e\frac{\partial M^z_{\text{orbit}}}{\partial \mu}
  =e\frac{\partial N}{\partial B^z_{\text{orbit}}},
 \label{eq:streda_formula}
\end{align}
the first equality of which is read off from the magnetization current $\bm{j}=\bm{\nabla}\times\bm{M}_{\text{orbit}}$ and an identity $e\bm{E}=-\nabla\mu$,
while the second equality is the Maxwell's relation of the free energy $F_{\text{em}}=\mu N-M^z_{\text{orbit}}B^z_{\text{orbit}}$.
The concept of the St\v{r}eda formula is inherited to the case of the thermal Hall effect\cite{nomura12,nakai16}, which leads to a crossed relation between thermodynamic quantities and gravitational quantities as\cite{nomura12}
\begin{align}
 \kappa_H 
  =\frac{\partial M^z_T}{\partial T}
  =\frac{\partial S}{\partial B_g^z}.
 \label{eq:streda_thermal}
\end{align}
Similar to the relation (\ref{eq:streda_formula}), the first equality of (\ref{eq:streda_thermal}) is the consequence of the heat magnetization current $\bm{j}_T=\bm{\nabla}\times\bm{M}_T$, where $M^z_T$ is the heat magnetization interacting with the gravitomagnetic field $B_g^z$ to define a free energy $F_{\text{th}}=-TS-M^z_TB_g^z$, and the second equality is the Maxwell's relation resulted from $F_{\text{th}}$.

In time-reversal symmetric topological insulators, spin-orbit coupling gives rise to fascinating interplay of the spin and the orbital magnetism\cite{yang06,murakami06,ito14,tserkovnyak15,koshino16}. 
The free energy of the angular momentum of electrons system with the spin and the orbital degrees of freedom is given by $F_{\text{mag}}= -\bm{M}\cdot\bm{B}=-(\bm{M}_{\text{orbit}}+\bm{M}_{\text{spin}})\cdot\bm{B}$.
The orbital angular momentum of the electron couples to the magnetic field through minimal coupling and is measured as the orbital magnetization as $M_{\text{orbit}}^a=-\partial F_{\text{mag}}/\partial B_{\text{orbit}}^a$, where $B_{\text{orbit}}$ is the intensity of the minimally coupled magnetic field and $a,b=x,y,z$.
On the other hand, the spin angular momentum of the electron and the magnetic field interact via Zeeman coupling, which is measured by the spin magnetization $M_{\text{spin}}^a=-\partial F_{\text{mag}}/\partial B_{\text{spin}}^a$, where $B_{\text{spin}}$ is the intensity of the magnetic field in the Zeeman coupling term.
Note that $B_{\text{orbit}}$ and $B_{\text{spin}}$ are distinguished by objects with which the magnetic field interacts, in order to extract the orbital and the spin magnetization separately, although both quantities are intrinsically same.
Besides the ordinary spin magnetic susceptibility $\chi^{ab}_{\text{spin}}=\partial M_{\text{spin}}^a/\partial B_{\text{spin}}^b$ and the orbital magnetic susceptibility $\chi^{ab}_{\text{orbit}}=\partial M_{\text{orbit}}^a/\partial B_{\text{orbit}}^b$, crossed magnetic susceptibilities are anticipated as the Maxwell's relation
\begin{align}
 \chi^{ab}_{\text{spin-orbit}}
 =
 \frac
 {\partial M_{\text{orbit}}^a}
 {\partial B_{\text{spin}}^b}
 =
 \frac
 {\partial M_{\text{spin}}^b}
 {\partial B_{\text{orbit}}^a}.
 \label{eq:soos}
\end{align}
Note that the crossed susceptibility is a part of the total magnetic susceptibility $\chi^{ab}=\chi^{ab}_{\text{spin}}+\chi^{ab}_{\text{orbit}}+2\chi^{ab}_{\text{spin-orbit}}$ induced by an external magnetic field interacting with both spin and orbital magnetism, and is specific to spin-orbit-coupled systems.
When the electron spin conserves $\sigma^z$, the crossed susceptibility $\chi^{zz}_{\text{spin-orbit}}$ is quantized in the quantum spin Hall insulators\cite{yang06,murakami06}, due to the following arguments.

A $\sigma^z$-conserved quantum spin Hall insulator consists of a combination of a quantum Hall insulator of spin-up electrons with the Chern number $\nu$ and that of spin-down electrons with the Chern number $-\nu$, where the spin Hall current is carried by the quantum Hall current of spin-up and spin-down electrons flowing in opposite directions.
In addition to current responses induced by the electric field, the quantum Hall system shows a quantized response to the magnetic field dictated by the St\v{r}eda formula $\sigma_H=\pm\nu e^2/h=e\partial N/\partial B^z_{\text{orbit}}$. 
The spin magnetization is then induced in the quantum spin Hall system by applying an external magnetic field, which increases and decreases the number of electrons according to relations $N_{\uparrow}=(\nu e/h)B^z_{\text{orbit}}$ and $N_{\downarrow}=-(\nu e/h) B^z_{\text{orbit}}$ and results in
\begin{align}
 M^z_{\text{spin}}
 =
 g\mu_{\text{B}}
 \left(
 N_{\uparrow}
 -
 N_{\downarrow}
 \right)/2
 =
 \frac{\nu e}
 {h} 
 g\mu_{\text{B}}
 B^z_{\text{orbit}}.
 \label{eq:spinmagnetization_qsh}
\end{align}

A counterpart of (\ref{eq:spinmagnetization_qsh}) from the relation (\ref{eq:soos}), that is, emergence of the orbital magnetization in the quantum spin Hall insulators induced by Zeeman coupling, can be understood as follows. 
Zeeman coupling raises and lowers the energy of spin-up and spin-down electrons by the same amount.
Thus the Zeeman coupling term behaves like a spin-dependent chemical potential term, which shifts the Fermi level of spin-up and spin-down electrons energy bands upward and downward or vice versa as $\mu_{\uparrow}=-\mu_{\downarrow}=g\mu_{\text{B}}B_{\text{spin}}^z/2$.
The shift of the Fermi level in the quantum Hall insulators gives rise to the emergent orbital magnetization according to the other form of the St\v{r}eda formula $\sigma_H=e\partial M_{\text{orbit}}^z/\partial \mu$.
The orbital magnetization in the quantum spin Hall system is then induced by Zeeman coupling by the relation
\begin{align}
 M_{\text{orbit}}^z
 =
 M_{\text{orbit},\uparrow}^z
 +
 M_{\text{orbit},\downarrow}^z
 =
 \frac{\nu e}{h}g\mu_{\text{B}}B_{\text{spin}}^z,
 \label{eq:orbitalmagnetization_qsh}
\end{align}
where the orbital magnetization of each $\sigma^z$-conserved electrons is $M_{\text{orbit},\uparrow}^z=(\nu e/h)\mu_{\uparrow}$ and $M_{\text{orbit},\downarrow}^z=-(\nu e/h)\mu_{\downarrow}$, respectively.
Two relations (\ref{eq:spinmagnetization_qsh}) and (\ref{eq:orbitalmagnetization_qsh}) are extensions of the St\v{r}eda formula to a time-reversal-symmetric case, as long as the electron spins are $\sigma^z$-conserved.

In this paper, we consider topological insulator models consisting of two spin-components and two orbital-components (conduction and valence bands) in two and three dimensions. 
We examine the orbital magnetization induced by two types of the Zeeman coupling terms, one is an ``orbital-independent'' Zeeman coupling term, which is proportional to the average of the g-factor of two orbitals, and the other one is an ``orbital-dependent'' Zeeman coupling term, which is proportional to the difference of two g-factors.
We also examine the crossed susceptibility even in the case of broken $\sigma^z$-conservation by estimating the crossed susceptibility as a function of the tilt angle of the magnetic field.
We then show a relation between the orbital magnetization in the bulk and the boundary persistent current induced by the magnetic field.
Furthermore, an argument based on symmetries is given to explain dependence of the orbital magnetization on g-factors.

This paper is organized as follows. 
In section \ref{sec:2d}, the crossed susceptibility of the Bernevig-Hughes-Zhang model of a two-dimensional quantum spin Hall system is estimated numerically. Its quantization feature is explained by the edge persistent charge current attributed to the chiral anomaly.
In section \ref{sec:3d}, the crossed susceptibility of the Wilson-Dirac model of a three-dimensional strong and weak topological insulators model is estimated and its behavior near phase transition points is explained using the surface Dirac fermion theory.
In section \ref{sec:discussion}, we prove that non-vanishing orbital magnetization requires breaking of both time-reversal symmetry and particle-hole symmetry. Also antisymmetric behavior of the crossed susceptibility as a function of the mass parameter is proven by another particle-hole type symmetry.

\section{Susceptibility of 2d topological insulator}
\label{sec:2d}

Consider the Bernevig-Hughes-Zhang (BHZ) model\cite{bernevig06} for the quantum spin Hall system,
\begin{align}
 H_{\text{BHZ}}(\bm{k})
 =
 \tau^x\sigma^z \sin k_x 
 +
 \tau^y \sin k_y 
 +
 \tau^z R_0(\bm{k}),
 \label{eq:bhz}
\end{align}
where $R_0(\bm{k})=m-2+\cos k_x+\cos k_y$.
The Pauli matrix $\sigma$ represents the ordinary electron spin, and $\tau$ represents the orbital degrees of freedom, where two components of the spinor correspond to wavefunctions of the s-orbital and that of the p-orbital.
The ground state of the BHZ model is $\mathbb{Z}_2$ nontrivial when $0<m<4$, and $\mathbb{Z}_2$ trivial otherwise.
Since the model conserves $\sigma^z$, each filled energy band is labeled by the eigenvalue of $\sigma^z$.
Within the $\mathbb{Z}_2$ nontrivial phase, there is a phase transition point at $m=2$ where the Chern number of each energy band is inverted from $\nu=\pm 1$ to $\nu=\mp 1$.
The Zeeman coupling term
\begin{align}
 U_{\text{Zeeman}}
 &=
 \begin{pmatrix}
  g_{1}\mu_{\text{B}}\bm{\sigma}\cdot\bm{B}/2 & 0 \\
  0 & g_{2}\mu_{\text{B}}\bm{\sigma}\cdot\bm{B}/2
 \end{pmatrix} \notag\\
 &=
 \begin{pmatrix}
  \bm{\sigma}\cdot(\bm{b}_++\bm{b}_-) & 0 \\
  0 & \bm{\sigma}\cdot(\bm{b}_+-\bm{b}_-)
 \end{pmatrix} \notag\\
 &=
 \tau^0\bm{\sigma}\cdot\bm{b}_+
 +
 \tau^z\bm{\sigma}\cdot\bm{b}_-
 \label{eq:zeeman}
\end{align}
is added to the BHZ model,
where $g_{1}$ ($g_{2}$) corresponds to the g-factor of the electron spin in the s- (p-) orbital\cite{liu10}.
Here we have defined two kinds of the magnetic field $\bm{b}_+=(g_{1}+g_{2})\mu_{\text{B}}\bm{B}/4$, which is referred to as the ``orbital-independent'' magnetic field, and $\bm{b}_-=(g_{1}-g_{2})\mu_{\text{B}}\bm{B}/4$, referred to as the ``orbital-dependent'' magnetic field, by associating and decomposing the Zeeman coupling terms for two orbitals.
While the BHZ Hamiltonian conserves time-reversal symmetry, the Zeeman coupling term breaks it, and thereby the orbital magnetization emerges.

\begin{figure}
 \centering
  \includegraphics[width=80mm]{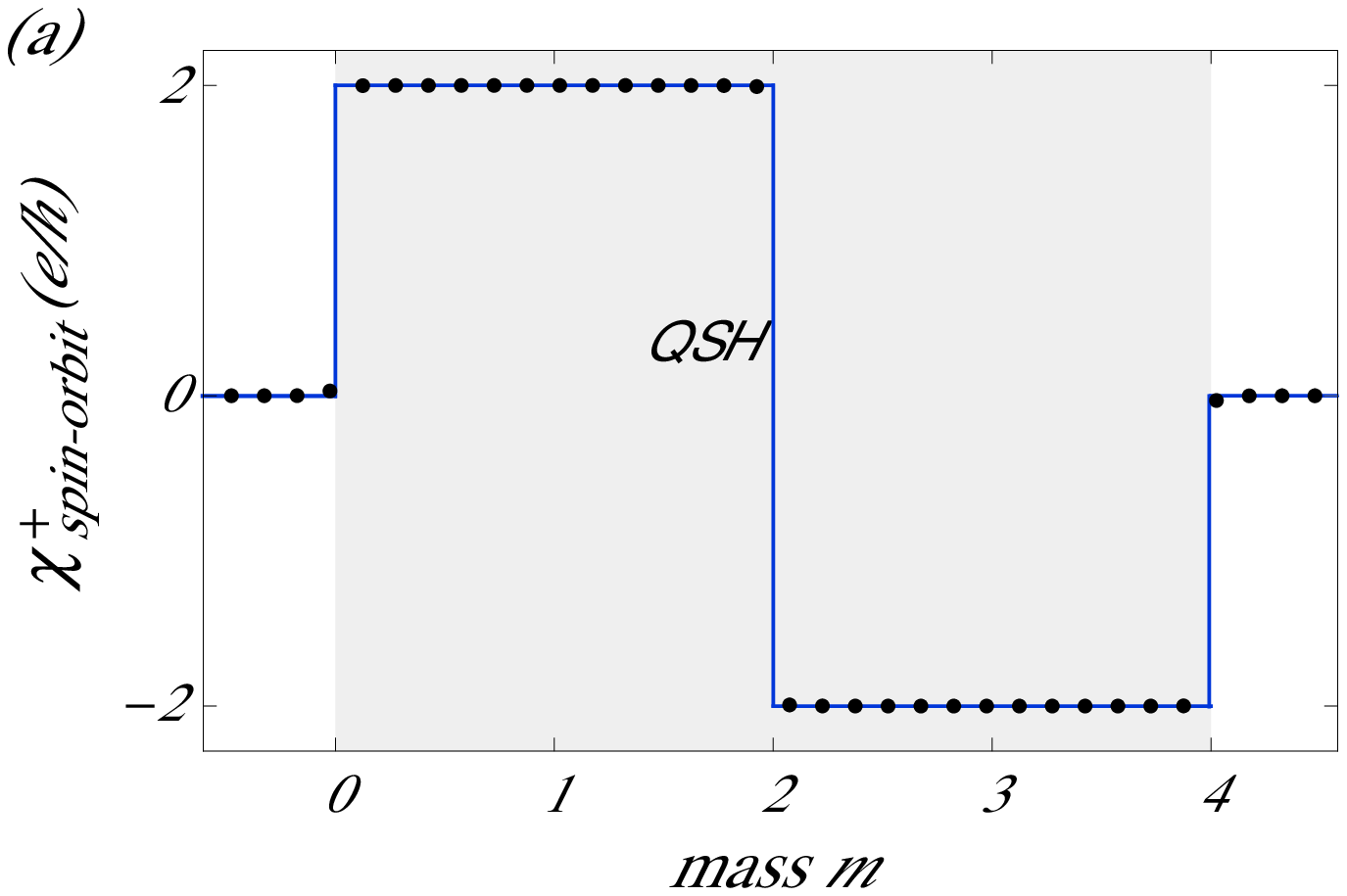}\\
  \includegraphics[width=80mm]{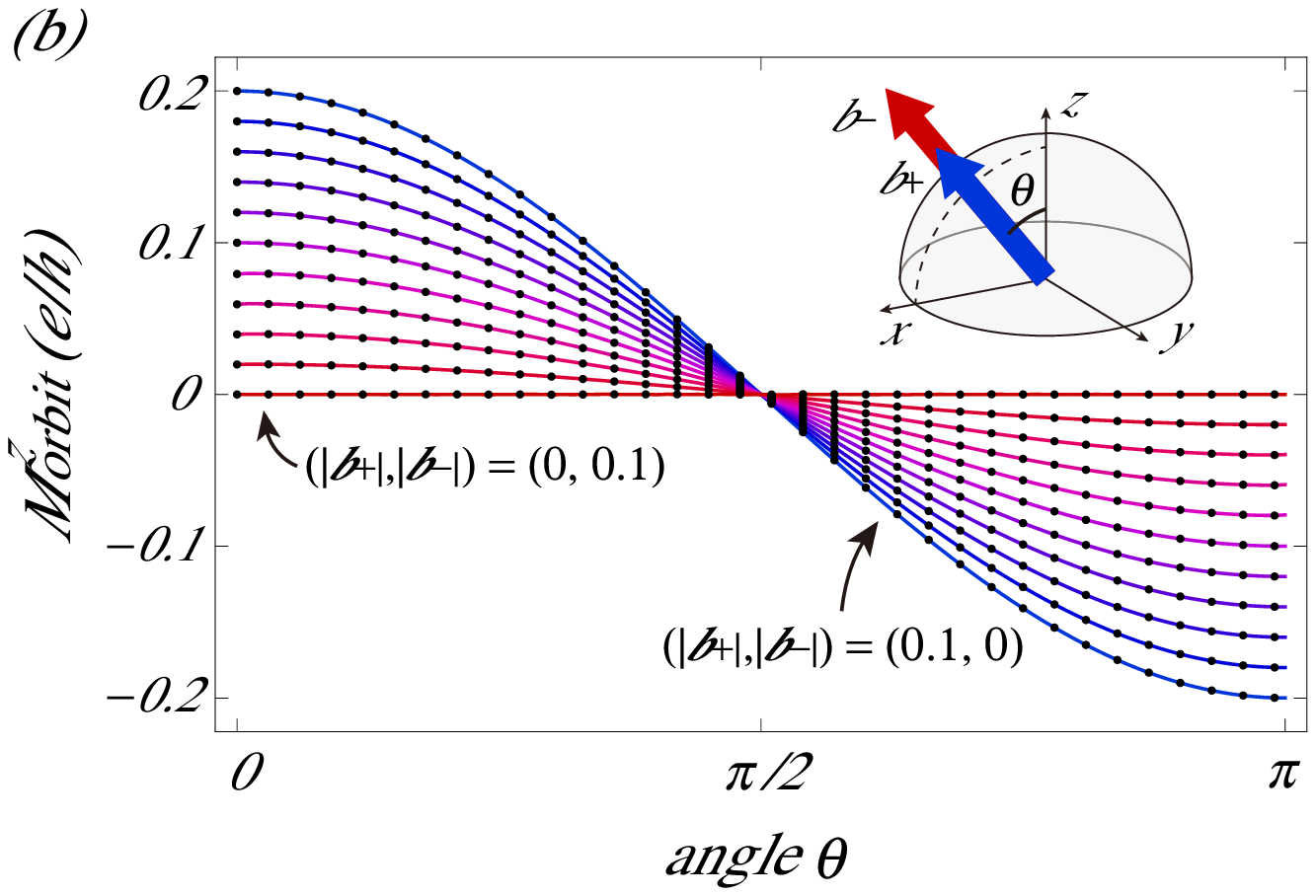}
 \caption{(Color online) (a) The crossed susceptibility $\chi_{\text{spin-orbit}}^{+}=\partial M_{\text{orbit}}^z/\partial b_+^z$ of the BHZ model is plotted as a function of the mass $m$. The solid line is the theoretically predicted value from (\ref{eq:orbitalmagnetization_qsh}) for the $\sigma^z$-conserved case, which is quantized at $\chi_{\text{spin-orbit}}^{+}=0,\pm2e/h$. (b) The orbital magnetization for a magnetic field tilted by the angle $\theta$ from the $z$-axis is plotted at $m=1$ for various value of the magnetic field $(|\bm{b}_+|,|\bm{b}_-|)=(0.1,0)$, $(0.09,0.01),\cdots,(0,0.1)$ keeping $|\bm{b}_+|+|\bm{b}_-|$ constant. \label{fig:mgnt2d}}
\end{figure}
The orbital magnetization of the system under the periodic boundary condition is formulated as\cite{sundaram99,xiao05,thonhauser05,ceresoli06,shi07}
\begin{align}
 \bm{M}_{\text{orbit}}
 &=
 \frac{e}{2\hbar}\sum_{n}
 \int \frac{d^2k}{(2\pi)^2} \notag\\
 &\quad\times
 \text{Im}
 \left\langle
 \frac{\partial u_{n,\bm{k}}}{\partial \bm{k}} 
 \right|
 \times
 (H(\bm{k})+E_{n,\bm{k}})
 \left|
 \frac{\partial u_{n,\bm{k}}}{\partial \bm{k}} 
 \right\rangle,
 \label{eq:orbital_magnetization}
\end{align}
where $E_{n,\bm{k}}$ and $u_{n,\bm{k}}$ are the eigenenergy and the Bloch wavefunction of the eigenstate of $H=H_{\text{BHZ}}+U_{\text{Zeeman}}$ labeled by the band index $n$ and the crystal momentum $\bm{k}$, and the summation of $n$ is taken over filled energy bands.
This quantity is numerically evaluated on the Brillouin zone defined by the region of $k_i\in[-\pi/a,\pi/a] (i=x,y)$, where $a$ is the lattice constant in the $x$ and the $y$ direction\cite{ceresoli06}.
The numerical calculation shows that the crossed susceptibility $\chi^{+}_{\text{spin-orbit}}=\partial M_{\text{orbit}}^z/\partial b_{+}^z$ induced by the magnetic field $b_+^z$ is quantized by $\pm 2e/h$ in the quantum spin Hall phase, and vanishes in trivial phases, both of whose results are proportional to the difference of the Chern number of two filled energy bands [Fig.~\ref{fig:mgnt2d}(a)].
The magnetization is evaluated when both magnetic field $\bm{b}_+$ and $\bm{b}_-$ are applied and the magnetic field is tilted from the $z$-axis within the $x$-$z$ plane by an angle $\theta$ [Fig.~\ref{fig:mgnt2d}(b)].
The result shows that the magnetization is proportional to $|\bm{b}_+|$ and $\cos\theta$ and independent of $|\bm{b}_-|$, which indicates that the orbital magnetization is sensitive only to $b_+^z=|\bm{b}_+|\cos\theta$.
The result also indicates that the quantization of the susceptibility $\chi^{+}_{\text{spin-orbit}}$ persists even when the $\sigma^z$ is not conserved, beyond the limited picture mentioned in (\ref{eq:spinmagnetization_qsh}) and (\ref{eq:orbitalmagnetization_qsh}).

The numerical results on the emergent orbital magnetization in the (2+1)-dimensional bulk
are consistent with the (1+1)-dimensional boundary theory.
There appear four time-reversal invariant momenta in the Brillouin zone where the occupied and unoccupied energy bands touch at the phase transition points. 
The energy dispersion can be linearized around these points when only the low-energy effective theory is concerned.
Conduction and valence energy bands touch at $\bm{k}=(0,0)$ when $m=0$, at $\bm{k}=(\pi,0),(0,\pi)$ when $m=2$, and at $\bm{k}=(\pi,\pi)$ when $m=4$.
Define the vacuum at $y>0$ by the Hamiltonian (\ref{eq:bhz}) with a large negative mass $M$ compared with the mass in the bulk of interest at $y<0$.
Note that the sign of the vacuum mass is irrelevant at the final result, since both limits of large mass correspond to the topologically trivial phase.
Gapless edge states appear when the sign of the mass changes at the boundary.
The low-energy edge Hamiltonian is given by (see Appendix~\ref{sec:edge})
\begin{align}
 H_{\text{edge}}
 =
 \sigma^z s p_x
 +
 \sigma^z b^z_+,
 \label{eq:le_edge_hamiltonian}
\end{align}
where $\bm{p}$ is a small crystal momentum measured from the center of the Dirac cone, and the sign $s$ is given by $s=-\text{sgn}(m)$ for the Dirac cone around $\bm{k}=(0,0)$, $s=\text{sgn}(m-2)$ for those around $\bm{k}=(\pi,0), (0,\pi)$, and $s=-\text{sgn}(m-4)$ for that around $\bm{k}=(\pi,\pi)$.
Therefore, while no edge state appears when $m<0$ and $m>4$, the boundary hosts single positive chirality edge mode when $0<m<2$, and single negative chirality edge mode when $2<m<4$.
Here pairs of opposite chirality edge modes are neglected.
It should be noted that none of the magnetic field except $b_+^z$ appears in the edge Hamiltonian (\ref{eq:le_edge_hamiltonian}), which agrees with the numerics.
The Zeeman coupling $b_+^z$ behaves like a spin-dependent chemical potential also in the edge theory.

The (1+1)-dimensional electrons system accommodates the chiral U(1) gauge anomaly\cite{fujikawa04, bertlmann96}, which brings about breaking of the conservation law of the chiral current $j_5^{\mu}=(e/\hbar)\langle\gamma^0\gamma^{\mu}\gamma_5\rangle$, where $\gamma^0=\sigma^x, \gamma^1=i\sigma^y, \gamma_5=\sigma^z$ in the model (\ref{eq:le_edge_hamiltonian}) with $s=1$.
The chiral anomaly is given by  
\begin{align}
 \partial_{\mu}j^{\mu}_5
 =
 -\frac{2e^2}{h}
 \epsilon^{\mu\nu}\partial_{\mu}A_{\nu},
 \label{eq:chiral_anomaly}
\end{align}
where $(eA_0,eA_1)=(0,b_+^z)$.
Assuming homogeneity of the magnetic field on the edge, integrals of the both side of (\ref{eq:chiral_anomaly}) over the perimeter of the edge space, the spatial derivative vanishes to give $\partial_0j_5^0=-(2e/h)\partial_0b_+^z$. Then adiabatically imposing the vector potential $A_1$ from zero to a finite value, one obtains
\begin{align}
 j_5^0
 =
 -\frac{2e}{h}b_+^z.
\end{align}
The relation of gamma matrix $\gamma^{\mu}\gamma_5=\epsilon^{\mu\nu}\gamma_{\nu}$ leads to an equality between the chiral charge current and the charge current $j^{\mu}=(e/\hbar)\langle\gamma^0\gamma^{\mu}\rangle$, given by $j^1=-j_5^0$, which is true for the (1+1)-dimensional Dirac electrons. 
The persistent edge charge current attributed to the chiral anomaly is thus $j^1=2eb^z_+/h$ for $s=1$, and similarly $j^1=-2eb^z_+/h$ for $s=-1$. 
Thus the total edge current is given by
\begin{align}
 j^1
 =
 \left\{
 \begin{array}{ll}
  0 \quad & (m<0 \text{ and } m>4)\\
  -2e b^z_+/h \quad & (0<m<2)\\
  2e b^z_+/h \quad & (2<m<4)
 \end{array}
 \right..
 \label{eq:edge_current}
\end{align}
Note that the charge current (\ref{eq:edge_current}) is a sort of the persistent current, which flows permanently under the magnetic field.
Persistent electric current in the quantum Hall system is accounted for by the magnetization current $\bm{j}=\bm{\nabla}\times\bm{M}_{\text{orbit}}$.
This relation connects the bulk orbital magnetization and the edge charge current by 
\begin{align}
 j^1
 =
 j^x
 =
 \int_{-\infty}^{\infty} \,dy \partial_y M_{\text{orbit}}^z(y) 
 =
 -M_{\text{orbit}}^z,
\end{align}
where the orbital magnetization in the bulk $M_{\text{orbit}}^z$ is defined by $M_{\text{orbit}}^z(y\to-\infty)$, and we used a fact that the orbital magnetization in the vacuum is zero.
Therefore the numerical result is qualitatively explained by the edge theory.

\section{Susceptibility of 3d topological insulator}
\label{sec:3d}

Next, we examine the crossed response in three-dimensional insulators, where none of intuitive argument like (\ref{eq:spinmagnetization_qsh}) and (\ref{eq:orbitalmagnetization_qsh}) exists.
We consider the Wilson-Dirac model of Bi$_2$Se$_3$-family three-dimensional topological insulators given by\cite{zhang09,liu10,wakatsuki15}
\begin{align}
 H_{\text{WD}}(\bm{k})
 =
 \sum_{i=x,y,z}
 \tau^x\sigma^i\sin k_i
 +
 \tau^z R_0(\bm{k}),
 \label{eq:3d}
\end{align}
where $R_0=m-3+\cos k_x+\cos k_y+\cos k_z$.
The Zeeman coupling term (\ref{eq:zeeman}) is added to this model.
In this model, two-components spinor diagonalizing the Pauli matrix $\tau^z$ represents bonding and antibonding orbitals of p$_z$ orbitals of Bi and Se.
Without the Zeeman coupling term, this model has the $\mathbb{Z}_2$ nontrivial phase when $0<m<2$ and $4<m<6$, and the $\mathbb{Z}_2$ trivial phase otherwise.
The ground state at $2<m<4$ is, however, weak-$\mathbb{Z}_2$ nontrivial (weak topological insulator phase).
Thus the surface Dirac fermions appear when $0<m<6$.

\begin{figure}
 \centering
 \includegraphics[width=80mm]{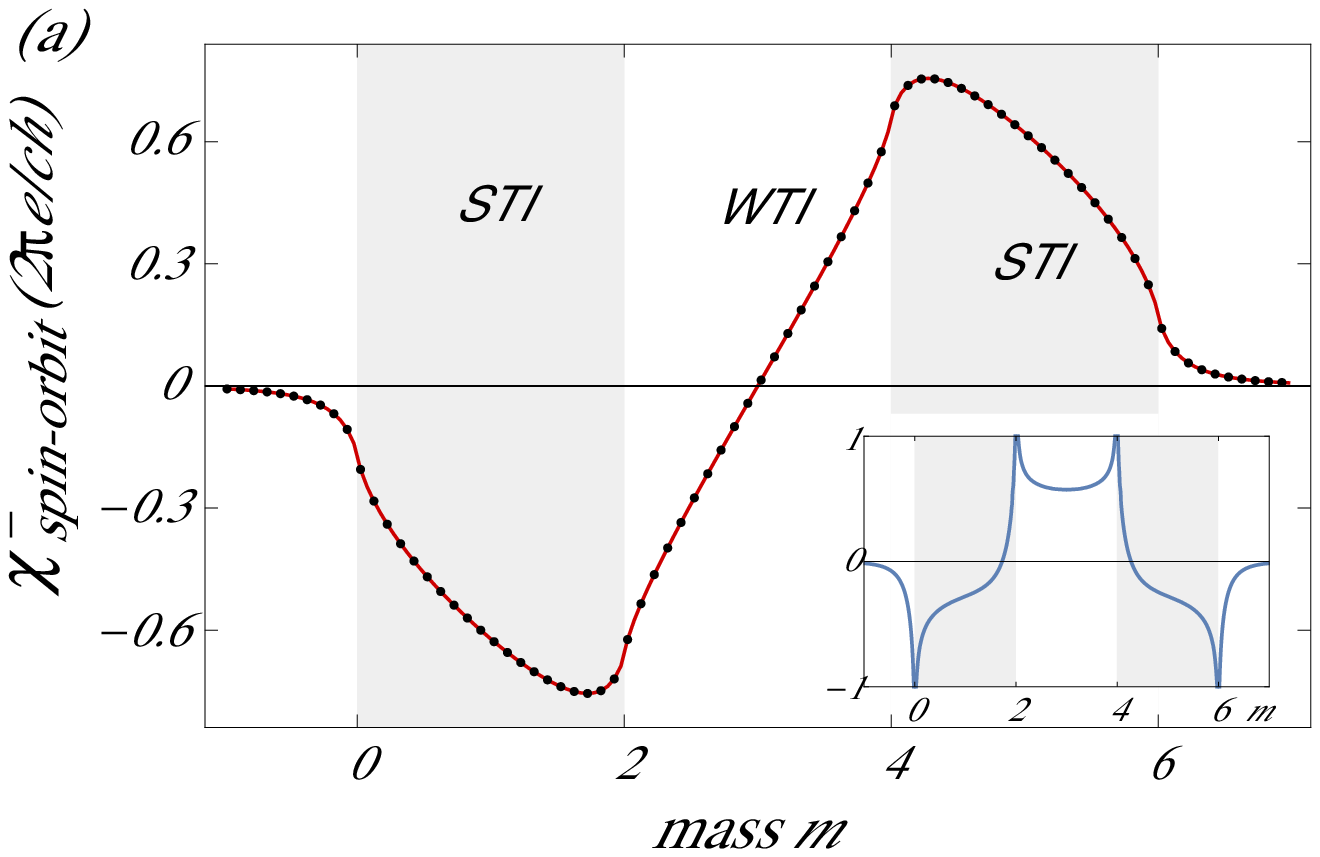} \\
 \includegraphics[width=82mm]{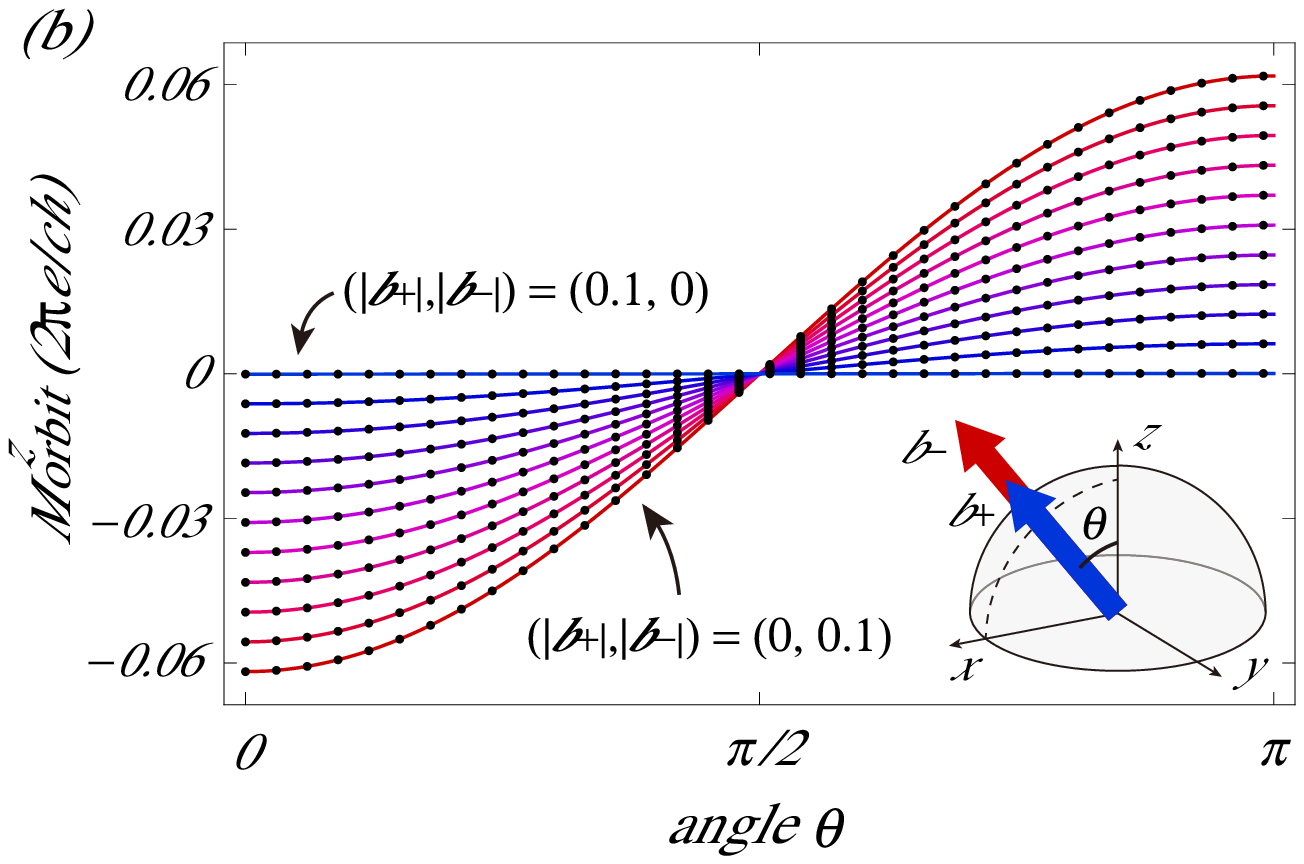}
 \caption{(Color online) (a) The crossed susceptibility $\chi_{\text{spin-orbit}}^-=\partial M_{\text{orbit}}^z/\partial b_-^z$ of the Wilson-Dirac model induced by the magnetic field $b_-^z$ is plotted as a function of the mass. The inset is the plot of its derivative $\partial \chi_{\text{spin-orbit}}^-/\partial m$. The shaded areas correspond to the strong topological insulator (STI) phase, while a non-shaded area in between corresponds to the weak topological insulator (WTI) phase. (b) The orbital magnetization induced by the magnetic field $\bm{b}_+$ and $\bm{b}_-$ with their tilt angle $\theta$ from the $z$-axis is estimated at $m=1$. Each plot corresponds to the external field $(|\bm{b}_+|,|\bm{b}_-|)=(0,0.1),(0.01,0.09),\cdots,(0.1,0)$ keeping the sum of them constant. \label{fig:mgnt3d}}
\end{figure}
The orbital magnetization is numerically evaluated by integrating the expression for the magnetization in two dimensions (\ref{eq:orbital_magnetization}) over $k_z$ from $-\pi/c$ to $\pi/c$, where $c$ is the lattice constant in the $z$-direction. [Fig.~\ref{fig:mgnt3d}(a)].
The crossed response coefficient $\chi_{\text{spin-orbit}}^-=\partial M_{\text{orbital}}^z/\partial b_-^z$ is finite around the region where the surface states appear, and is not quantized.
Even when both two types of the magnetic field $\bm{b}_+$, $\bm{b}_-$ exist and when the magnetic field is tilted from the $z$-axis, the orbital magnetization is sensitive only to the $z$-component of the orbital-\textit{dependent} magnetic field $b_-^z=|\bm{b}_-|\cos\theta$ [Fig.~\ref{fig:mgnt3d}(b)], which is in contrast to the two-dimensional topological insulator where the the orbital-\textit{independent} magnetic field induces the orbital magnetization.
Note that the crossed susceptibility at the phase transition point corresponds to that for three-dimensional Dirac semimetals.

Defining the vacuum by the Hamiltonian with large negative mass, the low-energy Hamiltonian of the surface states for the Dirac cone centered at $\bm{k}=(0,0,0)$ bound to the plane perpendicular to the $y$-axis is (see Appendix~\ref{sec:surface})
\begin{align}
  H_{\text{surf}}
 =&
 \sigma^z 
 \left(
 p_x
 +
 b_-^z
 \right) +
 \sigma^x
 \left(
 -p_z
 +
 b_-^x
 \right) 
 +
 \sigma^y
 b_+^y,
 \label{eq:le_surface_hamiltonian}
\end{align}
which appears when $m>0$.
Note that the surface Hamiltonian can be gapped by nonzero $b_+^y$ term, unlike the edge Hamiltonian (\ref{eq:le_edge_hamiltonian}).
There appears single right-handed Dirac fermion when $0<m<2$ and $4<m<6$, two left-handed Dirac fermions when $2<m<4$, and no Dirac fermion when $m<0$ and $m>6$.
Here pairs of Dirac fermions with opposite handedness are not considered, since such pairs can fuse to be gapped out by perturbations. 

The numerical results imply that the bulk magnetization in three-dimensional topological insulator cannot be derived from the purely (2+1)-dimensional linearized low-energy surface theory, since if the bulk magnetization is related to the low-energy surface theory, its value must be quantized in the region where the number of the surface Dirac cone is unchanged, that is, in the same topological phase, and also the magnetization must vanish once the surface state is gapped due to the in-plane magnetic field $b_+^y$.
The finite magnetization value tails which can be seen outside of the topological phase ($m<0$ and $m>6$) is considered to be from the higher energy valence bands.

Taking the above argument into account, the bulk magnetization is partly accounted for by the surface Dirac fermion theory accompanied with a precisely determined momentum cutoff.
The persistent surface current induced by the magnetic field at $0<m\ll 1$ is given by $\langle j_x \rangle\simeq-(m/2)^{1/2}e b_-^z/h$ (see Appendix \ref{sec:surface_current}), which explains sharp changes of the crossed susceptibility $\chi_{\text{spin-orbit}}^-$ appearing at the phase transition points $m=0$ and can be applied to points $m=2,4,6$ in Fig.~\ref{fig:mgnt3d}(a).
Such features are regarded as signature of appearance or disappearance of the Dirac cone at the surface, resulted from the phase transition between the strong topological insulator phase and the weak topological insulator phase or trivial phases.
Another momentum cutoff in the crystalline system naturally arises as the width of the Brillouin zone $k_i\in[-\pi/a,\pi/a]$ resulting from the lattice translational invariance. This type of cutoff may affect the magnetization away from the phase transition points. 
Similarly to the case of the boundary of the two-dimensional topological insulators, the surface current induced by the magnetic field is an persistent current, which contrasts to the inverse effect of the Edelstein effect\cite{edelstein90}.

Here we estimate a magnitude of the crossed susceptibility as compared with an ordinary magnetic susceptibility induced by Zeeman coupling, that is, the van Vleck susceptibility $\chi_{\text{spin}}=\partial M_{\text{spin}}^z/\partial H^z_{\text{spin}}$.
The van Vleck susceptibility of Bi$_2$Se$_3$ is numerically evaluated as\cite{yu10} $\chi_{\text{spin}} \sim 10^{-4}$.
The crossed susceptibility of the Wilson-Dirac model is of the order of $\chi^-_{\text{spin-orbit}}\sim g\mu_0\mu_{\text{B}}(2\pi e)/(ch)\sim 10^{-4}$, where $\mu_0$ is the permeability in the vacuum, and we have used parameters\cite{liu10} $g_1-g_2\sim 10$ and $c\sim 10\AA$.
This result indicates that the magnitude of the van Vleck and the crossed susceptibility are approximately of the same order.
Therefore, the crossed susceptibility provides an unignorable fraction among the total magnetic susceptibility in topological insulators.
However, at present, the van Vleck susceptibility, the crossed susceptibility and all other magnetic susceptibility cannot observed separately since the magnetic field equally interacts with the spin and the orbital magnetism.

\section{Discussion}
\label{sec:discussion}

Our results in previous sections have demonstrated that the orbital magnetization in the two-dimensional model is induced by the orbital-independent Zeeman coupling $\tau^0\sigma^zb_+^z$, while that in the three-dimensional model by the orbital-dependent Zeeman coupling $\tau^z\sigma^zb_-^z$.
In this section, we give a discussion on this property by proving that the orbital magnetization vanishes for (i) particle-hole symmetric Hamiltonians and (ii) time-reversal symmetric Hamiltonians.

At first, we consider particle-hole symmetry $\mathcal{C}^{-1}H(\bm{k})\mathcal{C}=-H(-\bm{k})$ which both two models implicitly possess when the chemical potential is tuned to be zero.
The charge conjugation operator for the BHZ model (\ref{eq:bhz}) is $\mathcal{C}_{\text{BHZ}}=\tau^x\sigma^z\mathcal{K}$, and that for the Wilson-Dirac model (\ref{eq:3d}) is $\mathcal{C}_{\text{WD}}=\tau^y\sigma^y\mathcal{K}$, where $\mathcal{K}$ is the complex conjugation operator.
The Zeeman coupling terms transform as follows,
\begin{align}
 &\mathcal{C}_{\text{BHZ}}^{-1}(\tau^0\sigma^z)\mathcal{C}_{\text{BHZ}}=\tau^0\sigma^z, \\
 &\mathcal{C}_{\text{BHZ}}^{-1}(\tau^z\sigma^z)\mathcal{C}_{\text{BHZ}}=-\tau^z\sigma^z, \\
 &\mathcal{C}_{\text{WD}}^{-1}(\tau^0\sigma^z)\mathcal{C}_{\text{WD}}=-\tau^0\sigma^z, \\
 &\mathcal{C}_{\text{WD}}^{-1}(\tau^z\sigma^z)\mathcal{C}_{\text{WD}}=\tau^z\sigma^z.
\end{align}
Thus the orbital-independent Zeeman coupling $\tau^0\sigma^zb^z_+$ breaks particle-hole symmetry of the BHZ model, while the orbital-dependent Zeeman coupling $\tau^z\sigma^zb^z_-$ breaks particle-hole symmetry of the Wilson-Dirac model.

When a Hamiltonian has particle-hole symmetry with respect to a charge conjugation operator $\mathcal{C}$, the orbital magnetization is shown to be zero due to the following argument.
Consider the integrand of the formula of the orbital magnetization (\ref{eq:orbital_magnetization}) given by 
\begin{align}
 &\bm{m}_{\text{orbit}}(\bm{k}) \notag\\
 &=
 \frac{e}{2\hbar}
 \sum_{n:\text{occ}}
 \text{Im}
 \left\langle
 \frac{\partial u_{n,\bm{k}}}{\partial \bm{k}} 
 \right|
 \times
 (H(\bm{k})+E_{n,\bm{k}})
 \left|
 \frac{\partial u_{n,\bm{k}}}{\partial \bm{k}} 
 \right\rangle \notag\\
 &=
 -\frac{e\hbar}{2}
 \sum_{n:\text{occ}}
 \sum_{m:\text{unocc}}
 \frac{E_{n,\bm{k}}+E_{m,\bm{k}}}{(E_{n,\bm{k}}-E_{m,\bm{k}})^2} \notag\\
 &\quad\cdot
 \text{Im}
 \left[
 \langle 
 u_{n,\bm{k}} 
 |\hat{\bm{v}}(\bm{k})|
 u_{m,\bm{k}}
 \rangle
 \times
 \langle
 u_{m,\bm{k}} 
 |\hat{\bm{v}}(\bm{k})|
 u_{n,\bm{k}}
 \rangle
 \right]
\end{align}
where ``occ'' and ``unocc'' indicate sets of indices of occupied and unoccupied energy bands, respectively.
Here the velocity operator is $\hat{v}_a(\bm{k})=\partial H(\bm{k})/\partial (\hbar k_a)$, and we have used a relation $\langle u_{n,\bm{k}}|\hat{v}_a|u_{m,\bm{k}}\rangle=(1/\hbar)(E_{n,\bm{k}}-E_{m,\bm{k}})\langle u_{n,\bm{k}}|\partial u_{m,\bm{k}}/\partial k_a\rangle$.
The velocity operator transforms under the charge conjugation as $\hat{v}_a(\bm{k})=\mathcal{C}^{-1}\hat{v}_a(-\bm{k})\mathcal{C}$.
Since the charge conjugation operator is antiunitary, inserting the unity operator $\mathcal{C}^{\dagger}\mathcal{C}=1$ between inner products gives
\begin{align}
 \langle
 u_{n,\bm{k}}
 |\hat{v}_a(\bm{k})|
 u_{m,\bm{k}}
 \rangle
 &=
 \langle
 u_{n,\bm{k}}|
 \mathcal{C}^{\dagger}\hat{v}_a(-\bm{k})\mathcal{C}
 |u_{m,\bm{k}}
 \rangle \notag\\
 &=
 \left(
 \langle
 u_{-n,-\bm{k}}
 |\hat{v}_a(-\bm{k})|
 u_{-m,-\bm{k}}
 \rangle
 \right)^{\ast} \notag\\
 &=
 \langle
 u_{-m,-\bm{k}}
 |\hat{v}_a(-\bm{k})|
 u_{-n,-\bm{k}}
 \rangle,
\end{align}
where $|u_{-n,-\bm{k}}\rangle=\mathcal{C}|u_{n,\bm{k}}\rangle$ is the eigenvector of the Hamiltonian $H(-\bm{k})$ with eigenvalue $E_{-n,-\bm{k}}=-E_{n,\bm{k}}$.
Using this equality, one obtains
\begin{align}
 &\bm{m}_{\text{orbit}}(\bm{k}) \notag\\
 &=
 -\frac{e\hbar}{2}
 \sum_{-n:\text{unocc}}
 \sum_{-m:\text{occ}}
 \frac{-E_{-n,-\bm{k}}-E_{-m,-\bm{k}}}{(E_{-n,-\bm{k}}-E_{-m,-\bm{k}})^2} \notag\\
 &\quad\cdot
 \text{Im}
 \left[
 \langle 
 u_{-m,-\bm{k}} 
 |\hat{\bm{v}}(-\bm{k})|
 u_{-n,-\bm{k}}
 \rangle
 \times
 \langle
 u_{-n,-\bm{k}} 
 |\hat{\bm{v}}(-\bm{k})|
 u_{-m,-\bm{k}}
 \rangle
 \right] \notag\\
 &=
 \frac{e\hbar}{2}
 \sum_{n:\text{occ}}
 \sum_{m:\text{unocc}}
 \frac{E_{n,-\bm{k}}+E_{m,-\bm{k}}}{(E_{n,-\bm{k}}-E_{m,-\bm{k}})^2} \notag\\
 &\quad\cdot
 \text{Im}
 \left[
 \langle 
 u_{n,-\bm{k}} 
 |\hat{\bm{v}}(-\bm{k})|
 u_{m,-\bm{k}}
 \rangle
 \times
 \langle
 u_{m,-\bm{k}} 
 |\hat{\bm{v}}(-\bm{k})|
 u_{n,-\bm{k}}
 \rangle
 \right] \notag\\
 &=
 -\bm{m}_{\text{orbit}}(-\bm{k}),
\end{align} 
which results in vanishing orbital magnetization for particle-hole symmetric Hamiltonians.
Thus the orbital-independent Zeeman coupling $\tau^0\sigma^zb^z_+$ induces the orbital magnetization in the BHZ model, while the orbital-dependent Zeeman coupling $\tau^z\sigma^zb^z_-$ does in the Wilson-Dirac model, which is consistent with the numerical results.

Similar kind of argument can be applied to prove vanishing orbital magnetization for time-reversal symmetric Hamiltonians.
Both the BHZ model and the Wilson-Dirac model has time-reversal symmetry with the time-reversal operator $\mathcal{T}=i\sigma^y\mathcal{K}$. 
Using $\hat{v}_a(\bm{k})=-\mathcal{T}^{-1}\hat{v}_a(-\bm{k})\mathcal{T}$, the matrix elements of the velocity operator has a property $\langle u_{n,\bm{k}}|\hat{v}_a(\bm{k})|u_{m,\bm{k}}\rangle=-\langle u_{\bar{m},-\bm{k}}|\hat{v}_a(\bm{k})|u_{\bar{n},-\bm{k}}\rangle$, where $|u_{\bar{n},-\bm{k}}\rangle=\mathcal{T}|u_{n,\bm{k}}\rangle$ is the time-reversal pair state of $|u_{n,\bm{k}}\rangle$ with the eigenenergy $E_{\bar{n},-\bm{k}}=E_{n,\bm{k}}$.
Then the orbital magnetization for time-reversal symmetric Hamiltonians follows a relation
\begin{align}
 &\bm{m}_{\text{orbit}}(\bm{k}) \notag\\
 &=
 -\frac{e\hbar}{2}
 \sum_{\bar{n}:\text{occ}}
 \sum_{\bar{m}:\text{unocc}}
 \frac{E_{\bar{n},-\bm{k}}+E_{\bar{m},-\bm{k}}}{(E_{\bar{n},-\bm{k}}-E_{\bar{n},-\bm{k}})^2} \notag\\
 &\quad\cdot
 \text{Im}
 \left[
 \langle 
 u_{\bar{m},-\bm{k}} 
 |\hat{\bm{v}}(-\bm{k})|
 u_{\bar{n},-\bm{k}}
 \rangle
 \times
 \langle
 u_{\bar{n},-\bm{k}} 
 |\hat{\bm{v}}(-\bm{k})|
 u_{\bar{m},-\bm{k}}
 \rangle
 \right] \notag\\
 &=
 \frac{e\hbar}{2}
 \sum_{n:\text{occ}}
 \sum_{m:\text{unocc}}
 \frac{E_{n,-\bm{k}}+E_{m,-\bm{k}}}{(E_{n,-\bm{k}}-E_{m,-\bm{k}})^2} \notag\\
 &\quad\cdot
 \text{Im}
 \left[
 \langle 
 u_{n,-\bm{k}} 
 |\hat{\bm{v}}(-\bm{k})|
 u_{m,-\bm{k}}
 \rangle
 \times
 \langle
 u_{m,-\bm{k}} 
 |\hat{\bm{v}}(-\bm{k})|
 u_{n,-\bm{k}}
 \rangle
 \right] \notag\\
 &=
 -\bm{m}_{\text{orbit}}(-\bm{k}),
\end{align}
which results in vanishing orbital magnetization.
Therefore, although the chemical potential term breaks particle-hole symmetry, the orbital magnetization is not induced even when the chemical potential term is added to the unperturbed Hamiltonian since it preserves time-reversal symmetry, while both types of the Zeeman coupling terms break time reversal symmetry:
\begin{align}
 \mathcal{T}^{-1}(\tau^0\sigma^z)\mathcal{T}=-\tau^0\sigma^z,\quad
 \mathcal{T}^{-1}(\tau^z\sigma^z)\mathcal{T}=-\tau^z\sigma^z.
\end{align}

A similar argument can also be applied to prove vanishing orbital magnetization at $m=3$ of the Wilson-Dirac model [Fig.~\ref{fig:mgnt3d}(a)].
The Hamiltonian (\ref{eq:3d}) at $m=3$ has accidental particle-hole symmetry $\mathcal{C}'=i\sigma^y\mathcal{K}'$, where $\mathcal{K}'$ is a complex conjugation operator accompanied with momentum reflection with respect to the point $\bm{k}_0=(\pi/2,\pi/2,\pi/2)$ which exchanges positions of the eight time-reversal invariant momenta as $(k_x,k_y,k_z)\to (\pi-k_x,\pi-k_y,\pi-k_z)$.
The Hamiltonian at $m=3$ respects this type of particle-hole symmetry, and the same is true for the Zeeman coupling term:
\begin{align}
 {\mathcal{C}'}^{-1}H_{\text{WD}}(\bm{k};m=3)\mathcal{C}'
 &=
 -H_{\text{WD}}(2\bm{k}_0-\bm{k};m=3), \\
 {\mathcal{C}'}^{-1}U_{\text{Zeeman}}\mathcal{C}'
 &=
 -U_{\text{Zeeman}}.
\end{align}
Since the velocity operator transforms as ${\mathcal{C}'}^{-1}\hat{v}_a(\bm{k})\mathcal{C}'=-\hat{v}_a(2\bm{k}_0-\bm{k})$, an equality $\bm{m}_{\text{orbit}}(\bm{k})=-\bm{m}_{\text{orbit}}(2\bm{k}_0-\bm{k})$ holds, which results in vanishing orbital magnetization at $m=3$.
Moreover, the charge conjugation operator $C'$ connects the Hamiltonian at $m$ with that at $6-m$ by 
\begin{align}
 &{\mathcal{C}'}^{-1}
 \left[
 H_{\text{WD}}(\bm{k};m)
 +
 U_{\text{Zeeman}}
 \right]
 \mathcal{C}' \notag\\
 &\quad=
 -H_{\text{WD}}(2\bm{k}_0-\bm{k};6-m)
 -
 U_{\text{Zeeman}}.
\end{align}
This property explains the numerical results  in Fig.~\ref{fig:mgnt3d}(a) where the susceptibility is anti-symmetric with respect to $m=3$.

\section{Conclusion}

In summary,
the crossed responses of the orbital magnetization induced by Zeeman coupling are estimated in time-reversal symmetric topological insulators in two and three dimensions.
By introducing two distinct g-factors for conduction and valence bands, we found that, according to presence or breakdown of particle-hole symmetry by Zeeman coupling, a qualitative difference of the crossed susceptibility between two- and three-dimensional cases appears, that is, the crossed susceptibility depends on an averaged g-factor in two dimensions while on difference of two g-factors in three dimensions.
In two-dimensional insulators, the susceptibility is quantized by the difference of the Chern number of two occupied energy bands in the quantum spin Hall phase, and the quantization of the susceptibility persists even when $\sigma^z$ is not conserved.
The orbital magnetization of the two-dimensional bulk is quantitatively explained by the persistent charge current of the (1+1)-dimensional low-energy edge theory attributed to the chiral anomaly.
In three-dimensional topological insulators, the orbital magnetization emerges only when the g-factors of conduction and valence bands are different.
The susceptibility is not quantized in the topological phases, and can be finite even in the trivial phase near topological phases.
From the surface theory, the magnetization of the bulk is partly explained by the cutoff-dependence of the surface persistent charge current, which can be regarded as a signature of appearance or disappearance of the surface Dirac cone at the quantum phase transition points.
From the symmetry viewpoint, we showed that Zeeman coupling with an averaged g-factor in the BHZ model and Zeeman coupling proportional to difference of two g-factors in the Wilson-Dirac model break particle-hole symmetry, which is essential in emergence of the orbital magnetization.
Also we showed that anti-symmetric behavior of the crossed susceptibility of the Wilson-Dirac model as a function of the mass is a consequence of another kind of particle-hole symmetry.

\acknowledgements
The authors acknowledge N.~Nagaosa for useful comments.
This work is supported by World Premier International Research Center
Initiative (WPI) and Grant-in-Aid for Scientific
Research (No.~15H05854 and No.~26400308) from MEXT,

\appendix

\section{Low-energy boundary Hamiltonian}

In this section, derivation of the low-energy edge Hamiltonian of the Bernevig-Hughes-Zhang (BHZ) model and the surface Hamiltonian of the Wilson-Dirac model is explained.

\subsection{Edge Hamiltonian of the Bernevig-Hughes-Zhang model}
\label{sec:edge}

The BHZ model has four time-reversal invariant momenta in the momentum space at $\bm{k}=(0,0), (0,\pi), (\pi,0)$, and $(\pi,\pi)$.
At first we consider the Dirac cone centered at $\bm{k}_0=(0,0)$.
The low-energy linearized Hamiltonian around $\bm{k}_0$ is given by
\begin{align}
 H(\bm{p})
 =
 \tau^y p_x 
 +
 \tau^x\sigma^z p_y
 +
 \tau^z m
 +
 \bm{\sigma}\cdot\bm{b}_+ 
 +
 \tau^z\bm{\sigma}\cdot\bm{b}_-,
 \label{eq:le_bulk_hamiltonian1}
\end{align}
where $\bm{k}=\bm{k}_0+\bm{p}$.
Here we consider the boundary at $y=0$ intervening the bulk at $y<0$ and the vacuum at $y>0$.
The vacuum is fictitiously expressed by the BHZ model with large mass limit $m\to \pm\infty$.
The boundary wavefunctions satisfy the equation
\begin{align}
 \left(
 -i\tau^x\sigma^z\partial_y
 +
 \tau^z m(y)
 \right)
 \psi_0
 =
 0.
 \label{eq:edge_equation}
\end{align}
and are given by
\begin{align}
 \psi_0
 =
 e^{ip_x x}
 \exp
 \left[
 -\tau^y\sigma^z
 \int_0^y
 dy'
 m(y')
 \right]
 |s\rangle.
\end{align}
The spinor $|s\rangle$ satisfies $\tau^y\sigma^z|s\rangle=-\text{sgn}(m)|s\rangle$, where $\text{sgn}(m)$ is the sign of the mass of the bulk, and spans a subspace of  the four-component spinor space.
Unit vectors of this subspace are 
\begin{align}
 |+,1\rangle
 =
 \frac{1}{\sqrt{2}}
 \begin{pmatrix}
  1 \\ i
 \end{pmatrix}
 \otimes
 \begin{pmatrix}
  1 \\ 0
 \end{pmatrix},\quad
 |+,2\rangle
 =
 \frac{1}{\sqrt{2}}
 \begin{pmatrix}
  1 \\ -i
 \end{pmatrix}
 \otimes
 \begin{pmatrix}
  0 \\ 1
 \end{pmatrix}
\end{align}
for $\tau^y\sigma^z|s\rangle=(+1)|s\rangle$, and
\begin{align}
 |-,1\rangle
 =
 \frac{1}{\sqrt{2}}
 \begin{pmatrix}
  1 \\ i
 \end{pmatrix}
 \otimes
 \begin{pmatrix}
  0 \\ 1
 \end{pmatrix},\quad
 |-,2\rangle
 =
 \frac{1}{\sqrt{2}}
 \begin{pmatrix}
  1 \\ -i
 \end{pmatrix}
 \otimes
 \begin{pmatrix}
  1 \\ 0
 \end{pmatrix}
\end{align}
for $\tau^y\sigma^z|s\rangle=(-1)|s\rangle$.
Constructing the two-component spinor $|\pm\rangle=(|\pm,1\rangle,|\pm,2\rangle)$, $4\times 4$ matrices composing the Hamiltonian are reduces to $2\times 2$ matrices acting on these spinors, as
\begin{align}
 &\tau^x\sigma^z, \tau^z, \sigma^x, \sigma^y, \tau^z\sigma^i (i=x,y,z) \to 0, \\
 &\tau^y\to -\text{sgn}(m)\sigma^z,
 \quad
 \sigma^z\to\sigma^z,
 \label{eq:projection_gamma_pauli_2d}
\end{align}
where the Pauli matrix in the right-hand side of (\ref{eq:projection_gamma_pauli_2d}) acts on the two-component spinor $|\pm\rangle$, while that on the left-hand side acts on the spinors $|\pm,1\rangle, |\pm,2\rangle$.
Then the edge Hamiltonian is given by
\begin{align}
 H_{\text{edge}}
 =
 -\text{sgn}(m)\sigma^z p_x
 +
 b^z_+\sigma^z.
 \label{eq:le_edge_hamiltonian1}
\end{align} 

The low-energy bulk Hamiltonian of the Dirac cone centered at $\bm{k}_0=(\pi,0)$ in terms of small deviation of the momentum defined by $\bm{p}=\bm{k}-\bm{k}_0$ is given from (\ref{eq:le_bulk_hamiltonian1}) by changing $p_x\to -p_x$, and $m\to m-2$, since $\sin (\pi+p_x)\simeq -p_x$, and $\cos (\pi+p_x)\simeq -1$. Therefore the low-energy edge Hamiltonian resulting from the Dirac cone at $\bm{k}_0=(0,\pi)$ is
\begin{align}
 H_{\text{edge}}
 =
 \text{sgn}(m-2)\sigma^z p_x
 +
 b^z_+\sigma^z.
 \label{eq:le_edge_hamiltonian2}
\end{align}
Here the boundary considered for (\ref{eq:le_edge_hamiltonian2}) is same as that for (\ref{eq:le_edge_hamiltonian1}).
The form of low-energy edge Hamiltonian for the Dirac cone at $\bm{k}_0=(0,\pi)$ is same as (\ref{eq:le_edge_hamiltonian2}), since $\sin (\pi+p_y)\simeq -p_y$ alters the sign in front of $\partial_y$ in (\ref{eq:edge_equation}) and the sign of the eigenvalue of the spinor as $\tau^y\sigma^z|s\rangle=\text{sgn}(m)|s\rangle$.
Similarly the low-energy edge Hamiltonian for the Dirac cone at $\bm{k}_0=(\pi,\pi)$ is
\begin{align}
 H_{\text{edge}}
 =
 -\text{sgn}(m-4)\sigma^z p_x
 +
 b^z_+\sigma^z.
\end{align}

\subsection{Surface Hamiltonian of the 3d topological insulator model}
\label{sec:surface}

The Wilson-Dirac model used in this study has eight time-reversal invariant momenta in the Brillouin zone specified by $k_i=0$ or $\pi$ for $i=x,y,z$.
The linearized form of the Hamiltonian of the Dirac cone centered at $\bm{k}=(0,0,0)$ is
\begin{align}
 H
 =
 \tau^x\bm{\sigma}\cdot\bm{p}
 +
 \tau^z m
 +
 \bm{\sigma}\cdot\bm{b}_+
 +
 \tau^z
 \bm{\sigma}\cdot\bm{b}_-.
\end{align}
Similarly to the case of the two-dimensional case,
we consider the surface perpendicular to the $y$-axis intervening the vacuum at $y>0$ and the bulk at $y<0$.
The surface state wavefunctions satisfy the equation
\begin{align}
 \left(
 -i
 \tau^x\sigma^y
 \partial_y
 +
 \tau^z m(y)
 \right)
 \psi_0
 =
 0,
\end{align}
and thus the subspace of the four-component spinor space of the surface mode is specified by $\tau^y\sigma^y|s\rangle=-\text{sgn}(m)|s\rangle$.
Then each matrix is projected as
\begin{align}
 &\tau^x\sigma^x\to\text{sgn}(m)\sigma^z,\quad
 \tau^x\sigma^z\to-\text{sgn}(m)\sigma^x,\notag\\
 &\sigma^y\to\sigma^y, \quad
 \tau^z\sigma^x\to\sigma^x, \quad
 \tau^z\sigma^z\to\sigma^z, \notag\\
 &\tau^x\sigma^y,\tau^z,\sigma^x,\sigma^y,\tau^z\sigma^y\to 0.
\end{align}
The surface Hamiltonian for the Dirac cone centered at $\bm{k}=(0,0,0)$ is then given by
\begin{align}
 H_{\text{surf}}
 =&
 \sigma^z 
 \left(
 sp_x
 +
 b_-^z
 \right) +
 \sigma^x
 \left(
 -sp_z
 +
 b_-^x
 \right) 
 +
 \sigma^y
 b_+^y.
\end{align}
where $s=\text{sgn}(m)$.
Using the same procedure, the surface Hamiltonian for the remaining Dirac cones are obtained.
In summary, the surface Hamiltonian is given by 
\begin{align}
 H_{\text{surf}}
 =&
 \sigma^x 
 \left(
 s_xp_x
 +
 b_-^z
 \right) +
 \sigma^y
 \left(
 s_zp_z
 +
 b_-^x
 \right) 
 +
 \sigma^z
 b_+^y.
\end{align}
with 
\begin{align*}
 (s_x,s_z)
 =\left\{
 \begin{array}{ll}
  (\text{sgn}(m),-\text{sgn}(m))\,\,&\text{for }\bm{k}=(0,0,0)\\
  (-\text{sgn}(m-2),-\text{sgn}(m-2))\,\,&\text{for }\bm{k}=(\pi,0,0)\\
  (-\text{sgn}(m-2),\text{sgn}(m-2))\,\,&\text{for }\bm{k}=(0,\pi,0)\\
  (\text{sgn}(m-2),\text{sgn}(m-2))\,\,&\text{for }\bm{k}=(0,0,\pi)\\
  (-\text{sgn}(m-4),-\text{sgn}(m-4))\,\,&\text{for }\bm{k}=(0,\pi,\pi)\\
  (-\text{sgn}(m-4),\text{sgn}(m-4))\,\,&\text{for }\bm{k}=(\pi,0,\pi)\\
  (\text{sgn}(m-4),\text{sgn}(m-4))\,\,&\text{for }\bm{k}=(\pi,\pi,0)\\
  (\text{sgn}(m-6),-\text{sgn}(m-6))\,\,&\text{for }\bm{k}=(\pi,\pi,\pi)\\
 \end{array}
 \right..
\end{align*}

\section{Surface current of three-dimensional topological insulators}
\label{sec:surface_current}

Consider a surface Hamiltonian of the $x$-$z$ plane resulted from the Dirac cone centered at $\bm{k}=(0,0,0)$ with $m>0$ given by
\begin{align}
  H_{\text{surf}}
 =&
 \sigma^z 
 \left(
 p_x
 +
 b_-^z
 \right) +
 \sigma^x
 \left(
 -p_z
 +
 b_-^x
 \right) 
 +
 \sigma^y
 b_+^y.
\end{align}
The eigenenergy of the occupied state is $E_{\bm{p}}=-\epsilon_{\bm{p}}=-[(p_x+b_-^z)^2+(p_z-b_-^x)^2+{b_+^y}^2]^{1/2}$ and corresponding eigenstate is
\begin{align}
 \phi_{\bm{p}}
 =
 \begin{pmatrix}
  p_x+b_-^z-\epsilon_{\bm{p}} \\
  -p_z+b_-^x+ib_+^y
 \end{pmatrix}
 /\sqrt{2\epsilon_{\bm{p}}(\epsilon_{\bm{p}}-p_x-b_-^z)}.
\end{align}
By expanding $R_0$ by terms up to quadratic in momenta near the Dirac point $\bm{k}_0=(0,0,0)$, the condition for appearance of the surface state is
\begin{align}
 R_0(\bm{k}_0+\bm{p})
 \simeq
 m-(p_x^2+p_z^2)/2
 >
 0.
 \label{eq:cutoff_condition}
\end{align}
Assuming the mass to be small enough, (\ref{eq:cutoff_condition}) is satisfied by momenta in a small circle $|\bm{p}|<\sqrt{2m}$, which gives the momentum cutoff.

The expectation value of the surface current is then
\begin{align}
 \langle j^x \rangle
 &=
 \frac{e}{\hbar}
 \int_{D_{\sqrt{2m}}}
 \frac{d^2 p}{(2\pi)^2}
 \phi_{\bm{p}}^{\dagger}\sigma^z\phi_{\bm{p}} \notag\\
 &=
 -\frac{e}{\hbar}
 \int_{D_{\sqrt{2m}}} \frac{d^2p}{(2\pi)^2}
 \frac{p_x+b_-^z}{\epsilon_{\bm{p}}},
 \label{eq:surface_current}
\end{align}
where $D_{\sqrt{2m}}$ is a circle of the radius $\sqrt{2m}$.
The expression (\ref{eq:surface_current}) shows that the expectation value of the surface current is essentially dependent on the cutoff momentum.
Expanding the surface current by the small magnetic field $b_-^z$, the zeroth order vanishes since the integrand is an odd function of the momenta.
The first order in $b_-^z$ is
\begin{align}
 \frac{d \langle j^x \rangle}{d b_-^z}
 &=
 -\frac{e}{\hbar}
 \int_{D_{\sqrt{2m}}} \frac{d^2p}{(2\pi)^2}
 \left(\frac{d}{d b_-^z}
 \frac{p_x+b_-^z}{\epsilon_{\bm{p}}}
 \right)_{b_-^z=0} \notag\\
 &=
 -\frac{e}{\hbar}
 \int_{D_{\sqrt{2m}}} \frac{d^2p}{(2\pi)^2}
 \frac{p_z^2}{|\bm{p}|^3} \notag\\
 &=
 -\frac{e}{\hbar}
 \int_{0}^{\sqrt{2m}} \frac{zdz}{4\pi}
 \frac{z^2}{z^3} \notag\\
 &=
 -\frac{e}{h}
 \sqrt{\frac{m}{2}}.
\end{align}
Thus for small magnetic field, the surface current flows according to
\begin{align}
 \langle j^x \rangle
 \simeq
 -\frac{e b_-^z}{h}
 \sqrt{\frac{m}{2}}.
\end{align}

\end{document}